\documentstyle[12pt]{article}
\parskip=\medskipamount
\def\be{\begin{equation}}
\def\ee{\end{equation}}

\newcounter{fig}

\begin{document}
\thispagestyle{empty}
\title{\bf Droplet Spreading:  Partial Wetting Regime Revisited}
\author{M.J.de Ruijter, J.De Coninck\footnote{Corresponding author} \\
Centre de Recherche en Mod\'elisation Mol\'eculaire,\\ 
Universit\'e de Mons-Hainaut, 20 Place du Parc, \\ 7000 Mons, Belgium
\and
and G.Oshanin \\
Laboratoire de Physique Th\'eorique des Liquides, \\
Universit\'e Pierre et Marie Curie, T.16, 4 place Jussieu, \\
75252 Paris Cedex 05, France}
\maketitle

\begin{abstract}
We study the time evolution of a sessile liquid droplet, 
which is initially put onto a solid surface in a non-equilibrium
configuration and then evolves towards its equilibrium shape.
We adapt here the standard approach to the dynamics of 
mechanical dissipative systems, in which the driving force,
i.e. the gradient of the system's Lagrangian function, is balanced
against the rate of the dissipation function. In our case    
the driving force is the loss of the droplet's free
energy due to the increase of its base radius, while
 the dissipation occurs due to  viscous flows in the core of the droplet 
and  due to frictional processes in the vicinity of the advancing 
contact line, associated with attachment of fluid particles to solid. 
Within this approach we derive closed-form equations for the 
evolution of the droplet's base radius, and specify several regimes
at which different dissipation channels dominate. 
Our analytical predictions compare very well with
experimental data.   
\end{abstract}
\pagebreak

\section{Introduction}
Many industrial and material processing operations 
require the spreading of a liquid on a
solid. To name but a few, we might mention coating and painting, plant
protection, glueing, oil recovery from porous rocks and lubrication.
Consequently, the liquid may be a paint, a lubricant, an ink or a dye.
The solid may either have a simple surface or be finely divided, as in
the case of suspensions, porous media or fibers [1-8].

Apart from the fundamental problem of whether a given solid is
wetted by the liquid in question,  many of the practical
applications require precise knowledge of the rates of the
wetting processes. Particularly, one is often interested to know
how fast a liquid droplet, deposited on a solid substrate, can
wet a given area of the solid surface.

In this regard, two different types of descriptions 
 of sessile droplet spreading exist: 
hydrodynamic and molecular kinetic, which differ from each other
mostly  in the consideration of the dominant dissipation channel
[1,3-8].
The first type of approach emphasizes the
dissipation due to viscous flows generated in the core of the spreading
droplet. Within this approach, which has been scrutinized 
by many authors during the last several decades [9-16],
a relation between the capillary number $Ca$ and the value of the contact
angle  $ \theta(t)$ has been derived. 
This relation, which represents a zeroth-order approximation 
in the expansion in powers of the capillary number,
 yields simple scaling laws determining the time evolution 
of the contact angle and of the droplet's base radius $R(t)$. 
 In case of spreading circular droplets, one finds
$R(t) \sim t^{1/10}$ and $\theta(t) \sim t^{- 3/10}$.  These
scaling laws have been examined experimentally and shown to  
agree fairly well with experimental data
for many liquid/solid systems [8-12,16,17].
On the other hand, already early 
 studies  by Sawicki \cite{sawicki}, 
who has examined experimentally  
spreading rates of liquid droplets composed of PDMS 
polymers having different molecular weights (and, respectively, different
viscosities), have demonstrated pronounced departures from
the $R(t) \sim t^{1/10}$ behavior for
 liquids of progressively lower and lower
viscosity.  Moreover, it has been realized (see e.g. Ref.4 for a general
discussion) that fitting experimental data 
to the hydrodynamic descriptions, (which presume 
certain cut-off at short 
length scales or at small contact angles),
often leads to unreasonably small values of these empirical 
cut-off parameters, which fall below
molecular size.

The second type of approach, which originates from
 the molecular kinetic theory of
Eyring \cite{eyring},  
has been  adapted to describe the kinetics of wetting
phenomena by
 Blake, 
Blake and Hayes \cite{blakec} and 
Cherry and Holmes
\cite{cherry}, 
and was subsequently developed in 
Refs.6,21 and 22.
In contrast to the hydrodynamic picture,
this approach
concentrates on the dissipative processes
 occuring in the vicinity of 
advancing contact
line, which stem from 
attachment of fluid particles to the solid, 
and ignores the dissipation
 due to viscous flows in the core 
of the liquid droplet. 
In this approach one finds 
for the dynamics of the base radius 
and contact angle 
$R(t) \sim t^{1/7}$ and $\theta(t) \sim t^{- 3/7}$,  respectively.  
These laws have also been shown 
to work fairly well for some
experimental liquid/solid
systems \cite{marmur,dodge,michela}, 
which contradicts apparently
the behavior reported in [8-12,16,17].
The origin of this discrepancy
 was under debate for a long time.

On the other hand, it has been clearly understood that both types of
dissipation do
 exist simultaneously and several  attempts to work out a combined
theory
have been made. In particular, a simple way of formulating a combined
theory
on the basis of the molecular kinetic approach was 
discussed in Ref.6.
It was supposed here that the
 effects of viscous flows can be
 incorporated into the
molecular kinetic 
approach directly, by simply 
adding to the barriers created by the
 liquid/solid attractions
a viscous contribution. This procedure results, however,
in equations which have essentially the same structure as those
originally found in Refs.18
 (apart from some renormalization of coefficients, which
now include the liquid's viscosity) and thus can not 
reproduce the experimentally seen $R(t) \sim t^{1/10}$
behavior.  Another way of thought has been put forward in the paper by 
 Voinov \cite{voinov}, who has 
conjectured an equation relating
the value of the 
microscopic cut-off angle $\theta_c$, which is artificially 
introduced in macroscopic
 hydrodynamic descriptions in 
order to remove
essential singularities, 
and non-hydrodynamic dissipation (hereafter denoted as $T\Sigma_l$). This 
idea
has been subsequently developed by 
Petrov
and Petrov \cite{petra},  
who have combined heuristically the 
relation between the 
 contact angle $\theta(t)$ and the capillary number $Ca$,
 obtained in terms of essentially
$\sl macroscopic$ hydrodynamic theory, 
the  Voinov's relation between the $\sl microscopic$ angle
$\theta_c$ and $T\Sigma_l$, 
and  the  Blake and Haynes expression
for the contact line velocity, which 
holds, in terms of this approximate approach,
 for any value of $Ca$.
Combination of these three relations 
allowed to derive a closed-form expression 
for the
contact line velocity as the function of
 the parameters of the liquid/solid system in
question. 
Evidently, the
resulting expression
can serve as a very useful  interpolation formula, 
but care should be exercised
in interpreting the fitting parameters as 
physical quantities since this effective approach is clearly
empirical. 
Lastly, a rather different
 approach has been put forward by de Gennes \cite{pgg} in his
analysis
of the energy dissipation in the
 precursor film. It was suggested in Ref.1
 that the unbalanced
capillary force should be compensated by
 the total energy dissipation
occuring during the
spreading process: namely, 
 the viscous dissipation in the core of the droplet,
dissipation at the advancing contact line and that in a precursor
film. 
In this analysis,
however, the emphasis was put on the latter
 dissipation channel and the
energy 
dissipation due to frictional processes in the vicinity of the
liquid/solid interface
 was intentionally neglected. 
On base of this analysis, it has been claimed afterwards in
\cite{broch} that it is most likely that 
the non-hydrodynamic dissipation dominates at
relatively large values of the contact angle, 
while the hydrodynamic dissipation channel
prevails at smaller angles. However, no selection of the 
kinetic regimes associated with
different dissipation channels
  has been performed  within this physically meaningfull
approach.

In the present paper we 
develop a macroscopic dynamic description
of a sessile liquid droplet  spreading on solid surface in a situation
appropriate
for partial wetting, i.e. such that the macroscopic spreading power
characterizing
the liquid/solid system in question is negative.
 In our approach we 
take into account two different  dissipation channels:
 dissipation due to viscous flows
and that due to frictional processes in the vicinity of
 the contact line. 
Dissipation in the precursor film, which is not a
generic feature
for partial wetting, is not considered here. To describe the time
evolution of
the droplet, we adapt the
standard mechanical approach to
 dissipative system dynamics \cite{lan}, in which the driving force -
the gradient of
the system's Lagrangian function is balanced by the
rate of total
dissipation. In our case the driving force is
the loss of the droplet's
free energy due to the increase in the base radius.
 Consequently, our approach complements that proposed by de
Gennes \cite{pgg}. Furthermore we derive closed-form 
equations describing
the time evolution of the droplet's base radius and discuss several
possible kinetic
regimes associated with different
 dissipation channels. We show that, as it was expected intuitively, 
both dissipation channels may be important, 
but they have a dominant effect on the kinetics of the wetting
process at different time scales: namely, 
non-hydrodynamic dissipation resulting in the law
$R(t) \sim t^{1/7}$ prevails at relatively short times, 
while hydrodynamic dissipation
which yields the $t^{1/10}$-law represents the dominant 
dissipation channel at long times.
Moreover, we find explicitly the crossover time separating
 these two regimes; we show
that it may be very large for liquids with low viscosity or substrates with high friction
coefficient,
 such that the $t^{1/10}$-law
will show up only at very long times, which is in
 agreement with the trend observed by
Sawicki \cite{sawicki} and Dodge \cite{dodge}.
This allows us to conclude that 
 those experiments which show the departure from
the $R(t) \sim
t^{1/10}$ behavior are apparently performed with liquid/solid systems with
the crossover time sufficiently large
compared to the time scale of observation. 
On the other hand, experiments which
reveal such a law are done with systems with low values of the crossover time.

Our paper is structured as follows. First, we formulate our model
and write down the
basic equations. Then, we propose a comparison between the
general analytical
equations, and experimental data. Afterwards, we 
analyse of different asymptotical regimes, which stem from the
different
dissipation
channels. Finally, we conclude with a
brief discussion of our results.

\section{The model and basic equations.}

Consider a non-volatile, sessile liquid droplet of volume $V$, 
which is placed  at time $t = 0$ on an ideal horizontal
solid substrate, exposed to a neutral gas phase, for example, air. 
We suppose that the initial shape of the
 droplet is part
of a sphere, characterized by the base radius $R_{0}$ and the macroscopic 
contact angle $\theta_{0}$ (see Fig.1). 
For simplicity, we will suppose in what follows that
$\theta_{0} \leq \pi/2$. 

The key parameter, which defines  the qualitative
behavior of the liquid droplet on a given substrate 
is the spreading power $S$. 
Explicitly, $S$
is given by
 \begin{equation} 
S \; = \; \gamma_{SG} \; - \; \gamma \; - \; \gamma_{SL},
\end{equation}
where $\gamma_{SG}$, $\gamma$ and $\gamma_{SL}$ denote the solid/gas,
liquid/gas and solid/liquid interfacial tensions respectively. 

When $S \geq 0$, the droplet spreads
spontaneously and tends to shield the solid from the gas phase.
 In the conventional 
macroscopic picture, the droplet spreads completely such that 
the final stage   
is a liquid layer, covering the solid surface.
More elaborate  theoretical descriptions have demonstrated, however,
that 
this is not necessarily so
and spreading may cease when the height at the apex of the droplet falls
to mesoscopic or microscopic scales  (see, e.g. Ref.1). At such
scales 
the disjoining pressure, which tends to thicken the droplet,
 becomes essential and competes effectively
against the capillary forces; in consequence, 
an equilibrium pancake like structure may arise 
instead of the molecular film. The thickness of this 
extended structure is determined by the interplay 
between the spreading power and the liquid/gas
interfacial tension $\gamma$. 

When $S < 0$, 
the liquid droplet partially wets the solid.
Depending on its initial shape,
 the macroscopic
droplet may contract or dilate but reaches
 eventually an equilibrium
spherical cap like shape. In the present work we will be
 concerned only with the latter case, supposing that 
the spreading power  is negative.
Consequently, in the situation under study, a liquid droplet, 
which is characterized initially by the
base radius $R_{0}$ and the contact angle $\theta_{0}$, will 
spread on the solid substrate until it reaches an
equilibrium  shape with base radius $R_{eq}$
 and contact angle $\theta_{eq}$ ($\theta_{eq} > 0$), which obeys the
Young's equation, written here as
\begin{equation}
cos\theta_{eq} \; = \; 1 \; + \; \frac{S}{\gamma}
\end{equation}

Further on, we will suppose that the droplet, which is 
placed initially in a non-equilibrium configuration,
retains the ideal spherical cap form at any moment in time. 
This implies that the instantaneous configuration
of the droplet can be totally described by a single parameter:
it is either the time-dependent base radius $R(t)$, the
contact angle $\theta(t)$, or the height at the apex of
 the droplet $h(t)$. Measuring, for instance, the
instantaneous  base radius, we can calculate the 
corresponding values of the dynamic contact 
angle by virtue of 
 the conservation of volume condition, which gives
\begin{equation}
\frac{R^{3}(t)}{V} \; = \; \frac{3}{\pi} \; \phi(\theta(t)),
\end{equation}
where
\begin{equation}
\phi(\theta(t)) \; = \;
\frac{(1 + \cos\theta(t))  \sin\theta(t)}{(1 - \cos\theta(t)) (2 +
\cos\theta(t))}
\end{equation}
The corresponding height at the apex of the droplet is then determined
by
\begin{equation}
h(t) \; = \; R(t) \; \frac{(1 - \cos\theta(t))}{\sin\theta(t)}
\end{equation}

We hasten to remark, that such an approximation is used here only for derivation 
simple explicit formulae.  Any other relationship between the contact
angle 
and the base radius could be incorporated within our approach, but would
not
significantly change the results.  Moreover, our approximation
is appropriate in its own right when
several reasonable physical conditions are fulfilled [25-28].
First, the droplet has to be 
sufficiently small, such that the Bond
number is small and thus gravity effects can be ignored. 
Second, attractive liquid-solid interactions, 
(say, van der Waals interactions), 
are to be sufficiently short-ranged. As a matter of fact, 
the interactions with the substrate will always distort the 
liquid edge in the vicinity of the
substrate. We suppose here that the typical size of the distorted 
region, which is normally microscopically
or mesoscopically large, is much smaller
than the characteristic scale of observation. Lastly, the 
viscosity of the liquid droplet 
has to be not very large. It is known, for instance, 
that for high-viscosity liquids, such as, liquids
composed of polymers with high molecular weight, only 
the upper part of the droplet can be
well approximated by a spherical cap; closer to the substrate 
there appears 
a macroscopically large protruded region - the so-called "foot".  
 
Within the spherical cap approximation, the potential 
energy of the gas/liquid/solid system  can be written 
as a function of $R(t)$ only. The associated free energy is then given
by
\begin{equation}
\begin{array}{c}
F\{R(t)\} \; =  \; \pi \; R^2(t) \; \Big(\gamma_{SL} - \gamma_{SG}\Big)
\; + 
2 \; \pi \; \gamma \; \int^{R(t)}_{0} \rho \; d\rho \;
\sqrt {1 \; + \; \Big(d
\xi(\rho)/d \rho\Big)^{2}}
\end{array}
\end{equation}
where the  terms under the integral describe, 
in the usual fashion, the 
surface energy of the droplet, while the other terms give the contributions due to
the  tensions of the  solid/gas and solid/liquid interfaces.
 The function $\xi(\rho)$ defines the height 
of the droplet at distance $\rho$ from the
symmetry axis (see Fig.1). From Eqs.(3) and (4), we 
find that for fixed $R(t)$ the local height 
$\xi(\rho)$  obeys
\begin{equation}
\xi(\rho) \; = \; \frac{R(t)}{\sin\theta(t)} \; \Big[ \Big(1 \; - \;
\frac{\rho^{2}
\sin^{2}\theta(t)}{R^{2}(t)}\Big)^{1/2}
\; - \; \cos\theta(t)\Big] 
\end{equation}

Let us now consider the spreading dynamics 
and address the question of how a droplet with initial radius $R_{0}$
evolves in time to the equilibrium state with
base radius $R_{eq}$.
To do this, we consider the droplet as a purely 
mechanical system, the configuration of which is uniquely defined 
by  Eq.(5) (in fact,
 by the instantaneous value of 
$R(t)$), 
and relate it
to the total dissipation occuring during
the change in the droplet radius from the value $R(t)$ to the value
$R(t) + 
\delta R(t)$, Ref.1.

Now, the relation between the driving force of spreading, derived from
Eq.(6),
 and the energy dissipation
is provided by 
the standard mechanical  description  of  dissipative systems dynamics
\cite{lan}.
Since $F\{R(t)\}$ is not explicitly dependent on time, 
the corresponding dynamic equation will contain only two terms,
\begin{equation}
\frac{\partial T\{R(t);\dot{R}(t)\}}{\partial \dot{R}(t)} \; = \; 
\; \frac{\partial F\{R(t)\}}{\partial R(t)},
\end{equation}
where $T\{R(t);\dot{R}(t)\}$ denotes the dissipation function, 
which describes
the total dissipation in a circular liquid droplet with base radius
$R(t)$ 
and a contact line moving with
velocity $\dot{R}(t)$, and the term on right-hand-side determines the
driving force of spreading - the 
derivative of the liquid/solid potential energy with
 respect to  the base radius. Actually, Eq.(8) 
is almost verbatim the
description proposed in Ref.1.

Substituting Eq.(7) into Eq.(6) and performing the integration under the
constraint of total volume
conservation, one obtains for the driving force on the right-hand-side of 
Eq.(8) the following
conventional expression
\begin{equation}
\frac{\partial F\{R(t)\}}{\partial R(t)} \; = \; - \; 2 \; \pi \; R(t) \;
\gamma \; (\cos\theta_{eq} \; - \; 
\cos\theta(t) )
\end{equation}
As a matter of fact, Eq.(9) insures that the right-hand-side of Eq.(8)
vanishes 
(and, consequently,
the spreading stops) when 
the droplet's base radius and the contact angle reach
 their equilibrium values, prescribed by Young's equation
and Eq.(3).

Now, following de Gennes \cite{pgg}, the dissipation function can be
represented as a sum
of three different components,
\begin{equation}
T\{R(t);\dot{R}(t)\} \; = \; T\Sigma_{l} \; + \; T\Sigma_{w} \; + \;
T\Sigma_{f},
\end{equation}
where the first term describes the dissipation occuring in the immediate 
vicinity of the contact line, the second one the losses due to viscous
flow in the core
of the droplet, while the third one stems from the dissipative processes
taking place in the precursor film.
In the complete wetting case, as shown in Ref.1, the third term is
of crucial importance - the entire
spreading power
$S$ is dissipated into the film.  In the partial
wetting case, however, the appearence of the precursor film is not
generic.
We therefore restrict our consideration to the partial wetting 
regime without a precursor, and do not consider any such a
dissipation here.  We also hasten to remark that Eq.(10) 
implicitly pressumes that
 the substrate is
rigid and cannot be
 deformed by the droplet, as may occur with such soft 
"solids" like human skin, rubber or elastomers. In this latter 
situation one should take into account an
additional dissipation channel which will result 
in interesting intermediate-time
behavior (see e.g. Refs.29 and 30). 
We will not consider such a possibility here. 

Let us discuss in more detail the forms of the two remaining terms,
namely, 
$T\Sigma_{l}$ and $T\Sigma_{w}$.
Dissipation in the vicinity of the contact line results from various
physicochemical processes which lead to
the attachment of liquid molecules to the solid. Such a dissipation 
channel was considered first in Refs.18 and 20, and
subsequently scrutinized  by several authors
(see, e.g. Refs.21,22 and references therein). 
According to Refs.18,  the
microscopic process associated with the 
dynamics of the contact line is
the hopping motion of the fluid molecules at the edge of the
droplet between
adsorption sites distributed on the solid surface; the motion of the
molecules 
at the contact line is not symmetric and they are displaced  
away from the droplet by molecules from the advancing liquid edge. 
At low velocities, the leading   contribution to $T\Sigma_{l}$
is:
\begin{equation}
T\Sigma_{l} \;  = \; 2 \; \pi \; R(t) \; \frac{\zeta_{0} \;
(\dot{R}(t))^{2}}{2},
\end{equation}
where $\zeta_{0}$ is the friction coefficient, which is
determined as \cite{blakec}:
\begin{equation}
\zeta_{0} \; = \; \frac{n \; k \; T}{K_{w}^{0} \; \lambda}
\end{equation} 
In Eq.(12), $T$ denotes the temperature, $k$ the Boltzmann constant,
 $n$ the concentration of
adsorption sites and
$K_{w}^{0}$ and
$\lambda$ are the typical jump frequency and length of
molecular displacements.  We note also
that Eq.(11) is a simple linearized
form 
of a more general result derived by Blake and Haynes \cite{blakec},
which strictly applies only for small capillary numbers
$Ca = \eta \dot{R}(t)/\gamma$, $Ca \ll 1$.

Next, the displacement of the contact line is followed (or preceded) 
by the redistribution of the fluid
particles in the core of the liquid droplet, which generates a 
complex flow pattern.
This flow pattern has been analysed in details both experimentally and
 theoretically (see, e.g. Refs.5,31 to 33
 and references therein). 
Experiments have evidenced a very characteristic
rolling motion, reminiscent of a caterpillar track; i.e. the liquid
 velocity at the free surface, which is
directed towards the advancing contact line, is larger than 
inside the bulk, and liquid particles that are
initially at the free surface ultimately move to the solid substrate. 
This rolling motion also gives rise to
viscous dissipation. 

To calculate the contribution of such flows to the 
total dissipation function, we make use of the
picture introduced by Seaver and Berg \cite{berg} in their
 derivation of the Voinov-Hofmann-Tanner law
$R(t) \sim t^{1/10}$. The results of this 
oversimplified
approach  differ from the rigorous analysis of
Cox \cite{cox}
only by insignificant numerical factors. Now, 
Seaver and Berg assumed that the fluid 
dynamics of the spreading
spherical cap can be approximated by that of a spreading cylindrical
 disk
of radius $R(t)$, height $h^{*}$ and volume equal to the volume $V$ of
 the droplet, see Fig.2. 
The height of the disk $h^{*}$ is not an independent parameter but is
fixed 
by the volume $V$  and  the radius $R(t)$,
\begin{equation}
h^{*} \; = \; \frac{R(t)}{3 \; \phi(\theta(t))}
\end{equation}   
Under the assumption that the
radial component $\it{V}_{\rho}$ of the fluid velocity  
is much greater than the $\xi$-component,
the continuity and $\rho$-momentum equations for the spreading
cylindrical disk reduce to
\begin{equation}
\frac{\partial (\rho \it{V}_{\rho})}{\partial \rho} \; = \; 0
\end{equation}
\begin{equation}
\eta \; [\frac{\partial}{\partial \rho} \; (\frac{1}{\rho} \; 
\frac{\partial (\rho \it{V}_{\rho})}{\partial \rho}) \; +
\; \frac{\partial^2 \it{V}_{\rho}}{\partial \xi^{\rm 2}}] \; = \; 0,
\end{equation}
where $\eta$ denotes the liquid viscosity. 

Solving Eqs.(14) and (15) subject to the no-slip boundary condition 
at $\xi = 0$, Seaver and Berg found that within the geometrical 
approximations involved and neglecting the
$\xi$-component of the flow velocity, the radial
flow in the spreading droplet can be viewed as 
quasi-steady laminar Couette flow
\begin{equation}
\it{V}_{\rho}(\rho,\xi) \; = \; A \; \frac{\xi}{\rho},
\end{equation}
where $A$ is a constant. To calculate this constant, Seaver and Berg
supposed that the radial shear stress at
$\xi = h^{*}$ was balanced by the effective radial surface tension,
which
yields eventually the desired
Voinov-Hofmann-Tanner law for the spreading droplet. To define the
dissipation in the core of the
droplet due to viscous flows,
we have to consider a different boundary condition at
the edge 
of the disk. Namely, we impose the condition 
that the upper part of the edge 
moves with a prescribed velocity $\dot{R}(t)$, i.e., $\it{V}_{\rho}(\rho
= R(t),\xi = h^{*}) =
\dot{R}(t)$, which gives for the $\rho$-component of the velocity
the following result:
\begin{equation}
\it{V}_{\rho}(\rho,\xi) \; = \; \frac{R(t) \; \xi}{h^{*} \; \rho} \;
\dot{R}(t)
\end{equation}
Now, the dissipation in the core of the droplet due only to radial
 flows is defined by
\begin{equation} 
T\Sigma_{w} \; = \; \eta \; \int_{V} dV (\frac{\partial
\it{V}_{\rho}}{\partial \xi})^{2}
\end{equation}
Substituting Eq.(17) into Eq.(18) and integrating over the droplet's
volume, we find
\begin{equation}
T\Sigma_{w} \; = \; 6 \; \pi \; R(t) \; \eta \; \phi(\theta(t)) \;
(\dot{R}(t))^{2} \; ln(R(t)/a),
\end{equation}
where the parameter $a$ 
denotes the lower cutoff value of 
$\rho$.  
Without this cutoff the core viscous dissipation will diverge, because 
the velocity 
 at the symmetry axis of the droplet has a singularity. This singularity 
is quite artificial, however.
Clearly, instead of the singular behavior predicted by Eq.(17)
one may expect that the radial velocity is  zero at the symmetry axis
and remains small within some cylindrical region of radius $a$ centered
at this axis. This is 
precisely the meaning of the parameter $a$ - it is the radius of the
core region, which is stagnant with
respect to the radial dilation. Here, we do not attempt to 
determine $a$ analytically, which would require the solution of complete
momentum and continuity equations, and will use it only as an adjustable
parameter, whose value will be extracted from the fit of the
experimental data. 
We note finally that as with Eq.(11), Eq.(19) 
determines only the leading contribution 
to the dissipation in the core of the droplet, appropriate  at low
capillary numbers $Ca$.

Combining Eqs.(19),(11),(10) and (9), we find from the balance of forces
given by Eq.(8) 
the desired  dynamic
equation, which describes the time evolution of the base radius and/or
of the
 time-dependent contact angle in the limit of low contact line
velocities:
\begin{equation}
\dot{R}(t) \; =  
\; \frac{\gamma \; \Big(\cos\theta_{eq} - \cos\theta(t)\Big)}{\Big[\zeta_{0} \; + \;
6
\; \eta \;
\phi(\theta(t)) \; ln(R(t)/a)\Big]}
\end{equation}
or, equivalently,
\begin{equation}
Ca \; =  \; \frac{\Big(\cos\theta_{eq} - \cos\theta(t)\Big)}{\Big[\delta \; + \; 2
\; 
 \phi(\theta(t)) \; ln(3
V \phi(\theta(t))/\pi a^3)\Big]},
\end{equation}
where the parameter $\delta$, which will be repeatedly used in what follows,
 is the ratio
of the friction coefficient for motion of the 
contact line on the solid
substrate and of the bulk viscosity, $\delta = \zeta_{0}/\eta$.

\section{Comparison of analytical results and experimental data.}

Ii has been demonstrated in  Ref.34
 that the results of both the molecular
kinetic
 and of  the hydrodynamic models alone can rather accurately fit
experimental
data. Since the equations evaluated 
here combine the results of both
models, we may expect 
that they would
fit experimental data at least equally as well
 as any of the above mentioned approaches or any other combined  approach
(see, e.g.  \cite{petra}).  
Here, however, one may claim
that  the
fitting parameters should 
be actually related to the physical characteristics of 
both the dissipation due
to
viscous flow in the core 
region of the droplet 
and to the dissipation near the contact line, 
since our approach is based on well-justified
physical grounds.

One of the systems studied in Ref.34 was a droplet 
of di-n-butylphthalate (DBP)  on 
poly(ethy-leneterephthalate) 
(PET) substrate.  At room temperature,
DBP
partially wets PET substrates with a very low 
equilibrium contact angle.  The viscosity and
surface
tension of DBP were 
respectively 19.6 mPa.s and 34.3 mN/m.  Here we use this set of
experimental data as a 
tentative test of the validity of the 
arguments presented above.  A more detailed
analysis of 
several experimental data sets will be published elsewhere.   

We fitted the current experimental 
 data by solving equation (21), which is a first order differential
equation in $R$ and $\theta$, using 
a fourth-order Runge-Kutta algorithm \cite{runge}.  Equations (3) and
(4) were used
to calculate the 
instantaneous value of the  base radius from the dynamic contact angle
(see Fig. 3).  
The fitted
parameters were $\theta_{eq}$, $\delta$ and 
$a$.  The errors between the fits and the experimental data were
calculated as follows:
\begin{equation}
E \; = \; \frac{1}{N} \; \Sigma_{i=1}^{N} \; |R_{i}^{c} - R_{i}^{m}|  \; 
\frac{100}{R_{0}},
\end{equation} 
 where N is the number of experimental data and $R_{i}^{c}$ and
$R_{i}^{m}$ are,
respectively, the calculated and 
measured base radius at time $t(i)$.  $R_{0}$ denotes the initial
measured base radius.
  The function $E$ can be seen as the average
error per measured datum 
point in per cent.
The result of the fitting is shown in Figure 3 and in
Table 1.  The $\theta_{eq}$, $\delta$ and $a$ for 
this excellent fit are respectively 0.1 degree, 130 and 1.4 $\mu$m.  The
average
error per measurement is less 
than 1 percent.  

Let us first consider the value of the  cutoff,
$a$.  It is much smaller than 
the dimensions of the droplet (mm), so that Eq.(17)
holds for most of the 
droplet.  On the other hand, $a$ is much larger than the molecular
length
(nm), which is consistent 
with the continuum description.  In conclusion, the
calculated value of $a$ supports 
the validity of the theory together with the assumptions underlying the
derivation of Eq.(20). 
The large value of $\delta$ indicates that the friction due to the bulk
flow is
much smaller than the 
friction in the vicinity of the contact line.  From $\delta$, the
friction
coefficient near the contact line 
$\zeta_{0}$ is 2.6 Pa.s.  Thus, the assumption of the hydrodynamic
approach,
namely that the dissipation 
near the contact line may be omitted, is not appropriate for the
liquid-solid
system under 
consideration.  This was also pointed out in Ref.34.

Next,  let us consider how the experimental data can be fitted by the
results
of the hydrodynamic approach only. 
If $\delta$ in Eq. (21) is forced to be 0, the equation becomes purely
hydrodynamic.  Under this 
condition, and using the same set of experimental data (see Fig. 4), we
find that  $a$ is approximately 
$0.044$ $\mu$m 
(Table 1).  To compensate for the dissipation near the contact line, $a$
has to be smaller than the 
value calculated before.  The fit is not as good as before, but the
modified equation still models 
the data very well (error is only about 1.2 percent).  

Let us consider the fit of the experimental data using the result
of the molecular kinetic model only:
We note Eq.(20) will reduce to the result of the 
linear capillary number
version of the Blake and Haynes
 molecular kinetic
theory, if we discard the second term in the denominator by setting, for
instance,
$\eta = 0$. 
The best fit under this condition is also shown in Figure 4. 
This time the fit is better except at very early times.  
The $\delta$ is now 330, indicating that by using only
the molecular kinetic 
theory, we  overestimate the friction near the contact line. 
 
In conclusion, by considering only the dissipation in the bulk
(hydrodynamic model), or only 
the dissipation near the contact line (molecular kinetic model), the
experimental data can still 
be fitted quite well.  However, the fitting parameters are adjusted to
compensate
for the omitted 
dissipation channels, and therefore loose some of their physical
meaning.

\section{Asymptotic solutions of the dynamical equations.}

{\bf Early-time dynamics.} Let us consider first the solution 
of Eq.(20) in the limit of short times, supposing, for simplicity, that
$\theta_{0} \approx \pi/2$. We find then from Eq.(20) 
that for sufficiently small deviations from the initial values of the
base radius and the contact angle, such that
\begin{equation}
R(t) \; - \; R_{0} \; \ll \; \frac{R_{0}}{ln(3 V/2 \pi a^{3})},
\end{equation}
\begin{equation}
\theta_{0} \; - \; \theta(t) \; \ll \frac{2}{ln(3 V/2 \pi a^{3})},
\end{equation}
these properties obey
\begin{equation}
R(t) \; = \; R_{0} \; + \; \frac{\gamma \; \cos\theta_{eq}}{ \zeta_{0}
\; + \; \eta} \; t
\end{equation}
and
\begin{equation}
\theta(t) \; = \; \theta_{0} \; - \; (\frac{16 \pi}{3 V})^{1/3} 
\; \frac{\gamma \; \cos\theta_{eq}}{ \zeta_{0} \; + \; \eta} \; t
\end{equation}
Substituting Eqs.(25) and (26) into Eqs.(23) and (24), we thus infer
that
this
 early-time linear dependence on
time persists until $t < t_{1}$, where
\begin{equation}
t_{1} \; \approx \; \frac{(\zeta_{0} \; + \; \eta) \; R_{0}}{\gamma \;
\cos\theta_{eq} \; ln(3 V/2 \pi a^{3})}
\end{equation} 

{\bf Intermediate-time dynamics.} At intermediate times, i.e. such
that $t \gg t_{1}$ but still much
less than the time at which $R(t)$ reaches its equilibrium value, 
it is not possible, of course, to obtain simple scaling laws for $R(t)$ 
of the form $R(t) \sim t^{\alpha}$. 
This becomes possible only when $\theta_{eq} \ll \pi/2$ and the contact
angle $\theta(t)$ 
gets sufficiently
small such that the expansion of the trigonometric functions in their
Taylor series 
up to the second power of $\theta(t)$ is justified. In this limit
Eq.(21)
simplifies considerably:
\begin{equation} 
 Ca  \Big[ 2  \delta  \theta(t)  +  \frac{16}{3}  ln(4
V/\pi a^{3} \theta(t))\Big]  
=  \theta(t)^{3}  -  \theta(t) \theta_{eq}^{2}
\end{equation}
In the limit $\delta \approx 0$, i.e. the limit when friction in the
vicinity
of the contact
line is negligibly small, Eq.(28) resembles, apart from the logarithmic
factors,
the result obtained earlier
by de Gennes
\cite{pggl} and Seaver and Berg \cite{berg}. 
Eq.(28) shows that in particular situations two different types of
scaling behavior
 can be observed  at this intermediate-time stage. 
For $\theta(t)$, which is small enough to insure expansion of the
trigonometric functions into the Taylor series, 
but still greater than certain $\theta_{c}$, which is approximately
defined as
\begin{equation}
\theta_{c} \; \approx \; \frac{8 \; ln(4 V/\pi a^{3})}{3 \; \delta},
\end{equation}
the first term in the brackets on the left-hand-side of Eq.(28) is the
dominant one. Consequently,
one should observe at this stage 
\begin{equation}
R(t) \; \sim \; (\frac{2 V}{ \pi})^{2/7} \; (\frac{14 \gamma
t}{\zeta_{0}})^{1/7},
\end{equation}
or, for the dynamic contact angle
\begin{equation}
\theta(t) \; \sim \; 2 \; (\frac{2 V}{\pi})^{1/7} \; 
(\frac{14 \gamma t}{\zeta_{0}})^{- 3/7},
\end{equation}
i.e. exactly the behavior predicted by the molecular kinetic approach
\cite{blakec,dodge}. 
Using Eq.(29) to (31) 
we can thus define the characteristic time until the scaling behavior
$R(t) \sim t^{1/7}$ is valid.
Substituting Eq.(31) into Eq.(29) we find
that Eqs.(30) and (31) can exist for times less than a certain
$t_{2}$, which equals
\begin{equation}
t_{2} \; \approx \; \frac{\zeta_{0} V^{1/3}}{2 \gamma} 
\; (\frac{\delta}{ln(V/a^{3})})^{7/3}
\end{equation}
This time can be sufficiently large, if the droplet's
 volume is large or $\delta$ is large. In this case,
 the regime described by
Eqs.(30) and (31) may persist over a wide time interval.

 At times greater than
$t_{2}$, but such that $R(t)$ is still
less than its equilibrium value, we will observe the crossover to
the Voinov-Hofmann-Tanner-type behavior
\begin{equation}
R(t) \; \sim \; (\frac{2 V}{\pi})^{3/10} \; 
(\frac{15 \gamma t}{\eta ln(3 V/\pi a^{3})})^{1/10},
\end{equation}
or, for the dynamical contact angle
\begin{equation}
\theta(t) \; \sim \; 2 \; (\frac{2 V}{\pi})^{1/10} 
\; (\frac{15 \gamma t}{\eta \; ln(3 V/\pi a^{3})})^{- 3/10}
\end{equation}
We note here that the crossover time $t_{2}$ appears to be a
fundamental parameter,
which distinguishes which of the dissipation chanels will dominate on
the scale of
experiments. Apparently, those experiments which show the departure from
the $R(t) \sim
t^{1/10}$ behavior are performed with liquid/solid systems with
characteristic time $t_{2}$ sufficiently large
compared to the time scale of observation. 
On the other hand, experiments which
reveal such a law are done with systems with low values of $t_{2}$.
In particular, 
the experimental data of Dodge \cite{dodge}, which favor the
molecular kinetic
prediction, i.e. $R(t) \sim
t^{1/7}$, were obtained for droplets composed of  PDMS-OH molecules.
These polymers
form strong chemical bonds with solid substrates and consequently one
can expect that
for such systems $\xi_{0}$ and $t_{2}$ are large, so that the
behavior predicted by Eq.(31) persists over an extended time interval.
We note
also that the
trend observed by Sawicki \cite{sawicki} is in complete qualitative 
agreement with our results;
namely, $t_{2}$, Eq.(32), depends strongly on the liquid's viscosity, 
$t_{2} \sim \eta^{-7/3}$, and thus can be very large for low-viscosity 
liquids.

{\bf Long-time relaxation to equilibrium.} Let us consider finally
the long-time behavior of the droplet
when it approaches the equilibrium shape. Representing
the base radius as $R(t) = R_{eq} - \delta R(t)$ and $\theta(t)$ as
$\theta(t) = \theta_{eq}
+ \delta \theta(t)$, where $\delta R(t)$ and $\delta \theta(t)$ denote 
small deviations\footnote{We note that $\delta R(t)$ and $\delta
\theta(t)$
are not independent but related to each other because of the 
volume conservation condition.} from the equilibrium values,
substituting  these expessions to Eq.(28) and
accounting for only linear terms, we have
\begin{equation}
\frac{\delta\dot{R}(t)}{\delta R(t)} \; \approx \; - 
\; \frac{3 \gamma \theta_{eq}^{2}}{\eta R_{eq} [\delta
\; + \; 8 ln(3 V/\pi a^{3} \theta_{eq})/3 \theta_{eq}]},
\end{equation}
which yields
\begin{equation}
\delta R(t) \; \approx \; \exp(- \frac{t}{T})
\end{equation}
with
\begin{equation}
T \; = \; \frac{\eta R_{eq} [\delta
\; + \; 8 \;  ln(3 V/\pi a^{3} \theta_{eq})/3 \theta_{eq}]}{3 \gamma
\theta_{eq}^{2}}
\end{equation}
We note that such an exponential form, with $T$ treated as an adjustable
parameter, has been
successfully used 
to fit experimental data in Ref.37. Our work 
thus provides an explicit interpretation of the characteristic time
$T$ in terms of the liquid/solid system parameters.

Let us now reconsider the experimental data presented above.  Using the
best values for 
$\theta_{eq}$, $a$, and $\delta$, we can extrapolate
 the spreading behaviour
of
our liquid/solid system at diffferent time scales.
  In Figure 5 we show the whole time range, from milliseconds up to
hours of 
  spreading, in a log-log plot, but within the constraints of $Ca < 1$
and $\theta < 90°$.  The calculated values for $t_{1}$,
$t_{2}$ and $T$
  are respectively $8.0 \times 10^{-3}$, $4.5$ and $1.5 \times 10^{4}$ 
seconds.  On
Figure 5 it is 
  shown that the time frame between $t_{1}$ and $t_{2}$ can best be
fitted by a straight line
  with slope 0.14, very close to the expected 1/7. Above $t_{2}$ the
curve is best
  fitted by a straight line with slope 0.10, which is exactly the
predicted value.  At
  times around T, the curve starts to deviate from this linear fit. 
This is an
  explanation for the different power laws which have been reported in
the literature. 
  When concentrating on systems 'far' from equilibrium, as is mostly the
case in 
  forced wetting experiments, the molecular kinetic theory should fit
the data very well.  
  On the other hand, when systems are studied close to equilibrium,
  we are likely to find the power law predicted by the hydrodynamic
models, or
  at very long times, to find an exponential behaviour.  
  
  Our experiments fall within the $10^{-2}$ - 10 seconds time frame,
around 
  the value of $t_{2}$.  This is exactly the reason why we could fit our
data almost equally 
  well with the hydrodynamic model as well as with the molecular kinetic
one.

\section{Concluding remarks.}

Using a combined dissipation channel approach, we obtained
 a macroscopic analytical description of droplet spreading on a solid
surface in
situations appropriate for partial wetting. We have shown that in the
general case, 
 a succession of several different regimes can be observed. Namely,
(1) a
fast early-time
stage characterized by a linear time-dependence of the base radius; (2)
a
kinetic
stage at which the
dominant contribution to dissipation comes from attachment of fluid
molecules to solid
and the time evolution of the base radius is compatible with the
predictions of the
molecular kinetic approches \cite{blakec};  (3) a kinetic stage at which
the
hydrodynamic
dissipation dominates and, lastly, (4) 
an exponential relaxation to the
equilibrium state.
Analysing experimental data along these lines allows us to
obtain meaningful 
values for the molecular friction and the necessary cut-off distance.

\section{Acknowledgements.}

The authors gratefully 
acknowledge 
fruitful and encouraging discussions
with Prof. A-M.Cazabat and  Dr. T.Blake. 
MdR wishes to thank Kodak European Research Laboratories for their 
hospitality during his visit 
and Dr. A.Clarke for his kind assistance in experimental 
measurements. This research was supported by the FNRS,
the European Community through the Grant CHRX-CT94-0448, the COST
Project D5/0003/95, PROCOPE-programme
 and by the Minist\`ere de la R\'egion Wallonne.

\pagebreak

\section{Figure Captions.}

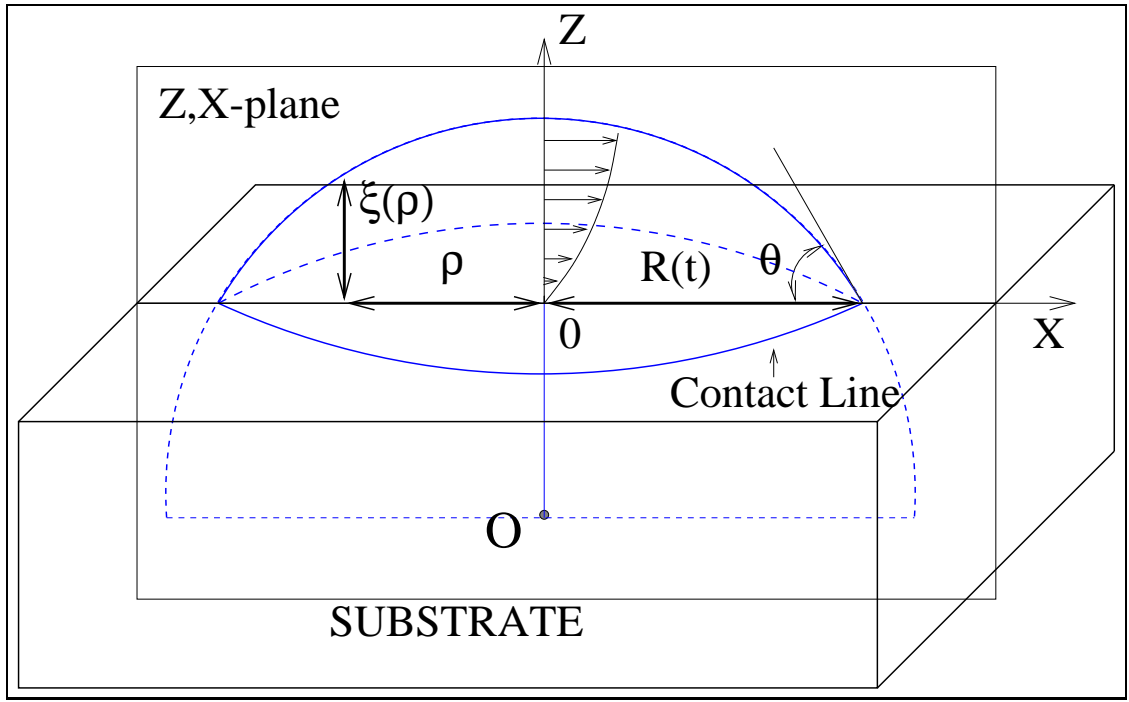
Figure 1. A schematic picture of a sessile droplet spreading on solid
substrate.

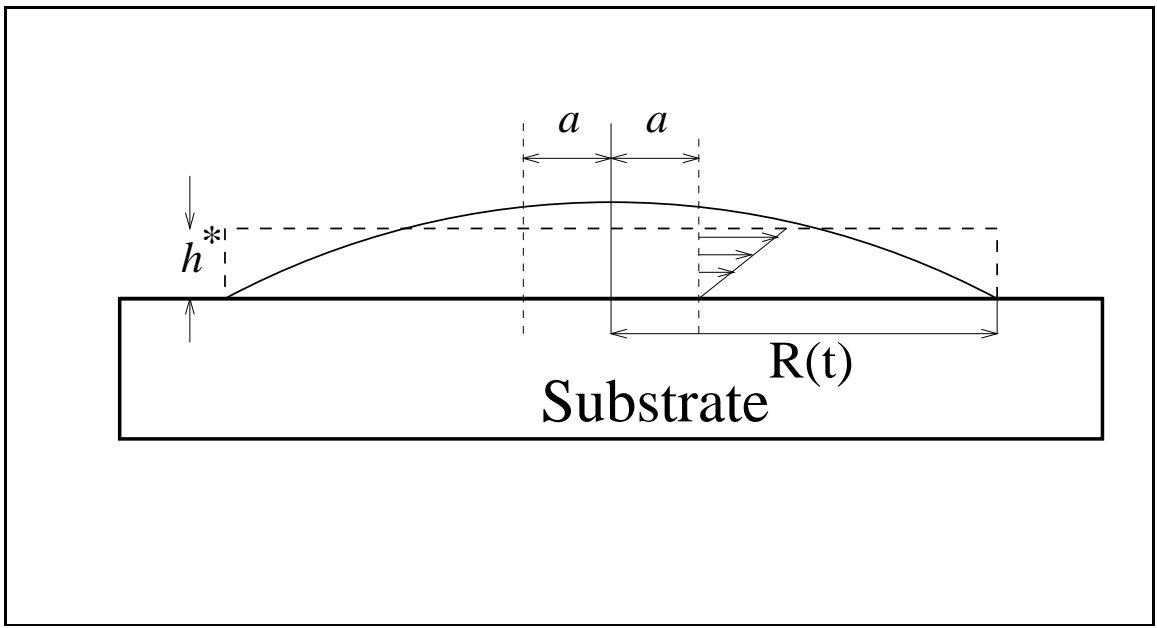
Figure 2. The Seaver-Berg  approximation of the flow pattern
in a sessile droplet spreading on a solid substrate.

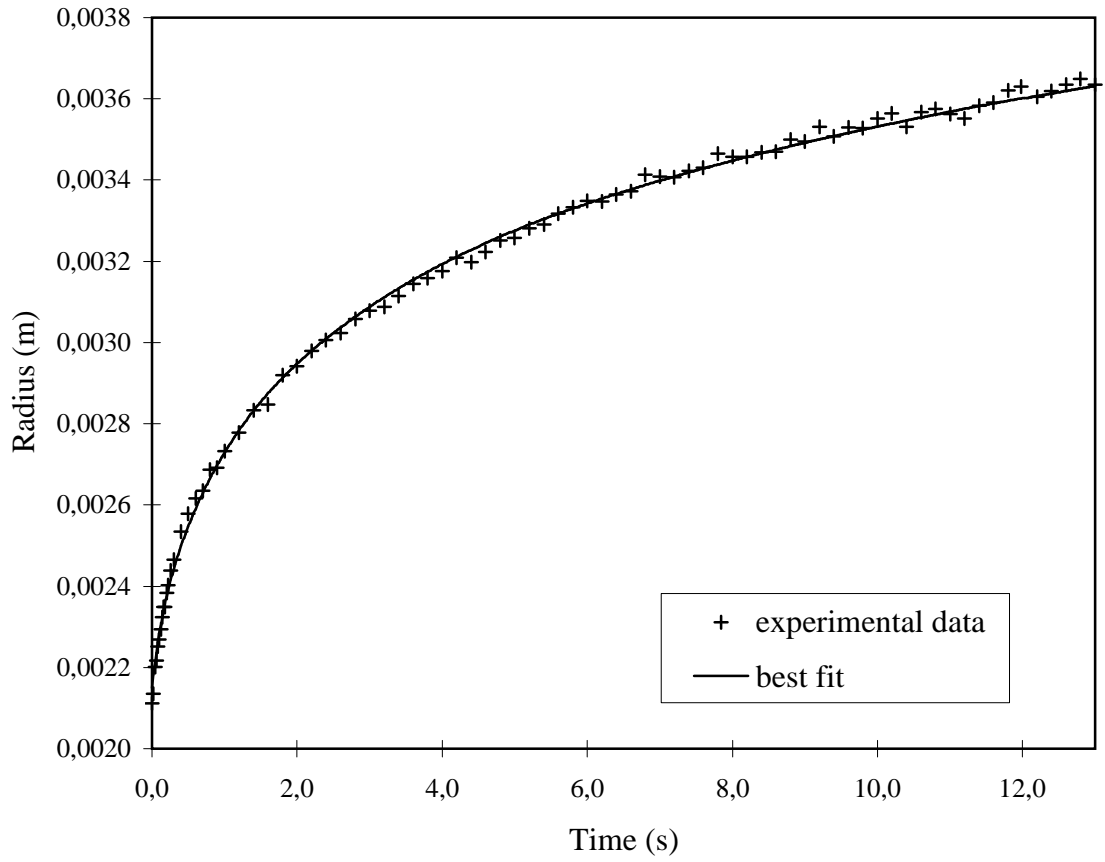
Figure 3.  The contact angle relaxation of DBP droplet on PET substrate
at room temperature.  The solid line is the
best fit, calculated with Eq. (20).  
The values of the fitting parameters are given in Table 1.

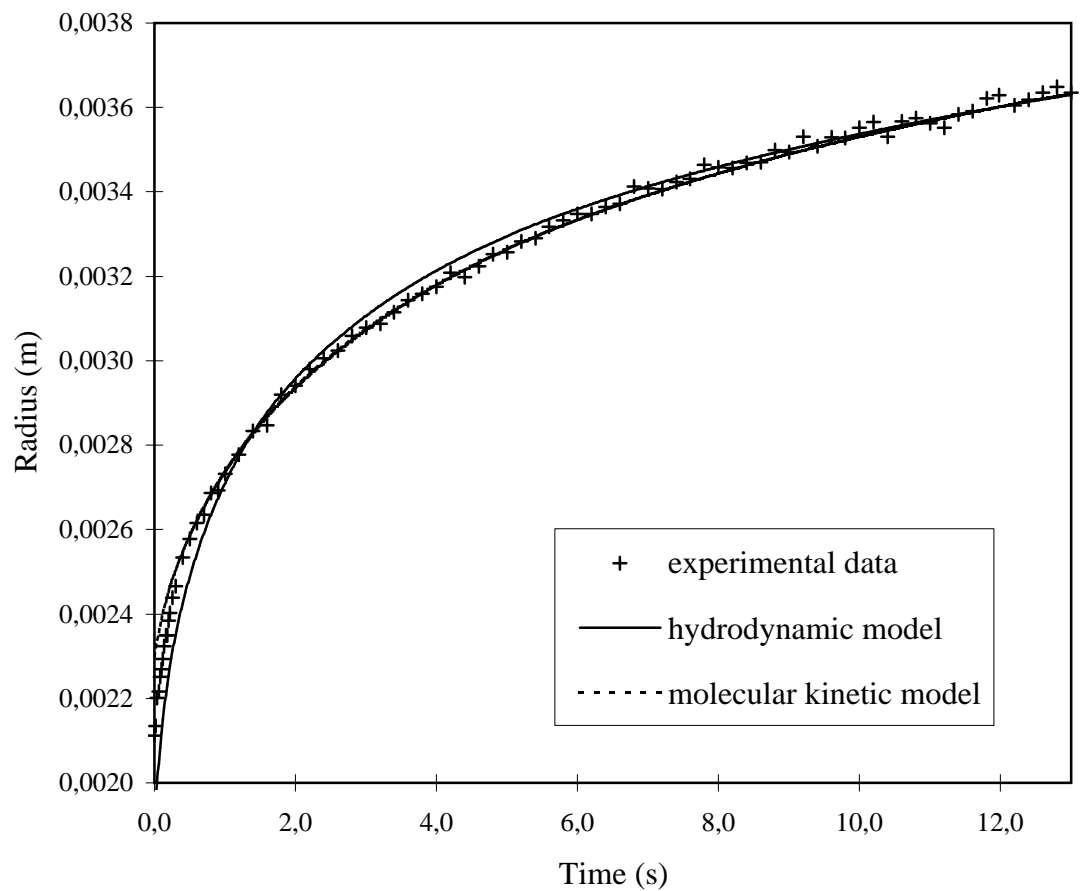
Figure 4.  The contact angle relaxation of DBP droplet on PET substrate
at room temperature.  The solid line
represents the best fit with the hydrodynamic model (Eq. (20) with
$\zeta_{0} = 0$).  
The dashed line is
the best fit with the molecular kinetic model (Eq. (20) with  $\eta =
0$).  
The best values for both
fits are given in Table 1.

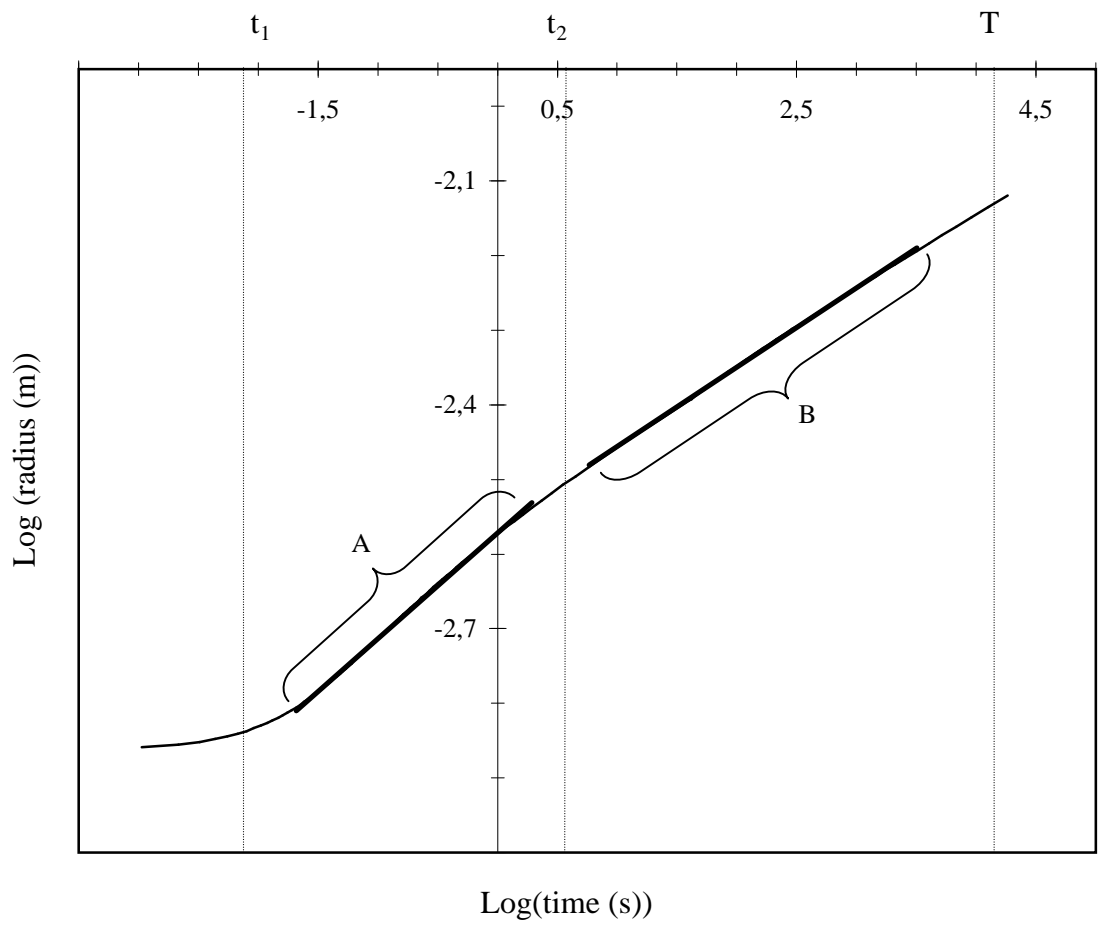
Figure 5.  Equation (20), with $\theta_{eq}  = 0.1$, $a = 1.4$,  
$\mu$ and $\delta = 130$.  The characteristic crossover times 
$t_{1}$, $t_{2}$ and $T$ are
calculated from Eqs.(27), (32) and (37).  Two
thick straight lines - $A$ and $B$ indicate the slopes $1/7$ and $1/10$,
respectively.  $A$: $Log(R) = 0,14  \; Log(t) - 2,57$, while $B$ is given by
$Log(R) = 0,10 \; Log(t) - 2,55$.

Table 1. Comparison of the fitting parameters using different models.  
The errors on $a$ and $\delta$ are 
calculated with the bootstrap method (Refs. 34,35), assuming a standard 
deviation of 1 degree on the
individual contact angle measurements.

\end{document}